\documentclass[conference,9pt]{IEEEtran}

\usepackage{algpseudocode}
\algtext*{EndWhile}
\algtext*{EndIf}
\algtext*{EndFor}
\algtext*{EndProcedure}
\usepackage{amsmath}
\usepackage{amssymb}
\usepackage{array}
\usepackage{algorithm}
\usepackage{bm}
\usepackage[bookmarks=false, hidelinks]{hyperref}
\usepackage[noadjust]{cite}
\usepackage[font=small]{caption}
\usepackage{color}
\usepackage{comment}
\usepackage{dblfloatfix}
\usepackage[inline]{enumitem}
\usepackage{epsfig}
\usepackage{etoolbox}
\usepackage{fancybox}
\usepackage{float}
\usepackage{fixltx2e}
\usepackage{graphicx}
\usepackage{listings}
\usepackage{mathrsfs}
\usepackage{multirow}
\usepackage{multicol}
\usepackage{pifont}
\usepackage{placeins}
\usepackage{rotating}
\usepackage{setspace}
\usepackage{soul}
\usepackage[hang, tight]{subfigure}
\usepackage{threeparttable}
\usepackage{times}
\usepackage{url}
\usepackage{upgreek}
\usepackage{verbatim}
\usepackage{wrapfig}
\usepackage[usenames,dvipsnames]{xcolor}
\usepackage{authblk}
\usepackage[]{siunitx}
\usepackage[super]{nth}
\usepackage{tabularx}
\usepackage{supertabular}
\usepackage{tikz}
\usepackage{bbm}
\usepackage{blindtext}

\graphicspath{{./_fig/}}

\usepackage{bbding}

\makeatletter
\renewcommand\footnoterule{%
  \kern-3\p@
  \hrule\@width0.4\columnwidth
  \kern2.6\p@}
  \makeatother

\usepackage{tikz}
\usepackage{xcolor}

\usepackage[hang,flushmargin]{footmisc}

\begin{document}

\title{AutoPower: \underline{Auto}mated Few-Shot Architecture-Level \underline{Power} Modeling by Power Group Decoupling\vspace{-.11in}}


\author[]{ \fontsize{11}{11}\selectfont Qijun Zhang, Yao Lu, Mengming Li, Zhiyao Xie\textsuperscript{*}\vspace{-6pt}}

\affil[]{\fontsize{10}{10}\selectfont Hong Kong University of Science and Technology\vspace{-7pt}}

\affil[]{	\{qzhangcs, yludf, mengming.li\}@connect.ust.hk, eezhiyao@ust.hk\vspace{-12pt}}

\maketitle
\begingroup\renewcommand\thefootnote{*}
\footnotetext{Corresponding Author}
\endgroup

\begin{abstract}

Power efficiency is a critical design objective in modern CPU design. Architects need a fast yet accurate architecture-level power evaluation tool to perform early-stage power estimation. However, traditional analytical architecture-level power models are inaccurate. The recently proposed machine learning (ML)-based architecture-level power model requires sufficient data from known configurations for training, making it unrealistic.

In this work, we propose AutoPower targeting fully automated architecture-level power modeling with limited known design configurations. We have two key observations: (1) The clock and SRAM dominate the power consumption of the processor, and (2) The clock and SRAM power correlate with structural information available at the architecture level. Based on these two observations, we propose the power group decoupling in AutoPower. First, AutoPower decouples across power groups to build individual power models for each group. Second, AutoPower designs power models by further decoupling the model into multiple sub-models within each power group. 
In our experiments, AutoPower can achieve a low mean absolute percentage error (MAPE) of 4.36\% and a high $R^2$ of 0.96 even with only two known configurations for training. This is 5\% lower in MAPE and 0.09 higher in $R^2$ compared with McPAT-Calib, the representative ML-based power model.
    
\end{abstract}

\section{Introduction}

Power efficiency is a critical design objective in modern CPU design, and its optimization relies on efficient power evaluation. However, as the complexity of processors keeps scaling up, standard VLSI power evaluation flow requires increasing time. The standard power evaluation flow goes through RTL implementation, RTL simulation, logic synthesis, and power simulation~\cite{ptpx}. Completing such a power evaluation flow can take weeks, seriously hindering early microarchitecture optimization and design space exploration. Architects need a fast yet accurate architecture-level power evaluation to support the early optimization of CPU microarchitecture.

\textbf{Analytical Power Model:}  Classical architecture-level power models such as McPAT~\cite{li2009mcpat} and Wattch~\cite{brooks2000wattch} are analytical models~\cite{zhang2023panda} designed by engineers for a specific type of processor\footnote{Engineers carefully analyze the hardware behavior of each component when relevant events occur to infer the activity of the hardware circuit, and then utilize their own experience and experimental data to estimate the power consumption of the component when relevant events occur.}. However, as processor architecture keeps updating, the analytical power model can hardly reflect the design of new processors. 
As mentioned in many studies~\cite{xi2015quantifying, zhai2022mcpat,zhang2023panda}, directly applying McPAT on the new generation of CPUs incurs unacceptable errors~\cite{zhang2023panda}. Therefore, it is usually necessary to modify McPAT for each target architecture, requiring huge engineering effort and time overhead.

\textbf{Machine learning (ML)-based Power Models:} In recent years, ML-based architecture-level power models~\cite{zhai2022mcpat,zhai2023microarchitecture} are trained to predict the CPU power consumption based on hardware parameters and event parameters. However, training an accurate ML-based model \emph{requires a large number of samples with golden power labels}. In practice, it is time-consuming to go through a standard power evaluation flow to collect golden power labels. More importantly, the number of available configurations is also limited.
Therefore, these data-hungry ML-based models are not applicable in real engineering scenarios. Some recent ML works~\cite{zhang2023panda} proposed to reduce required training data by unifying analytical models and ML-based models. However, \cite{zhang2023panda} relies on analytical resource functions, which are design-dependent and heavily based on architect expertise. Therefore, there is still a lack of a fully automatic power modeling method that only requires limited known design configurations for training.

\begin{figure}[!t]
\centering
\vspace{-.1in}
\includegraphics[width=0.45\textwidth]{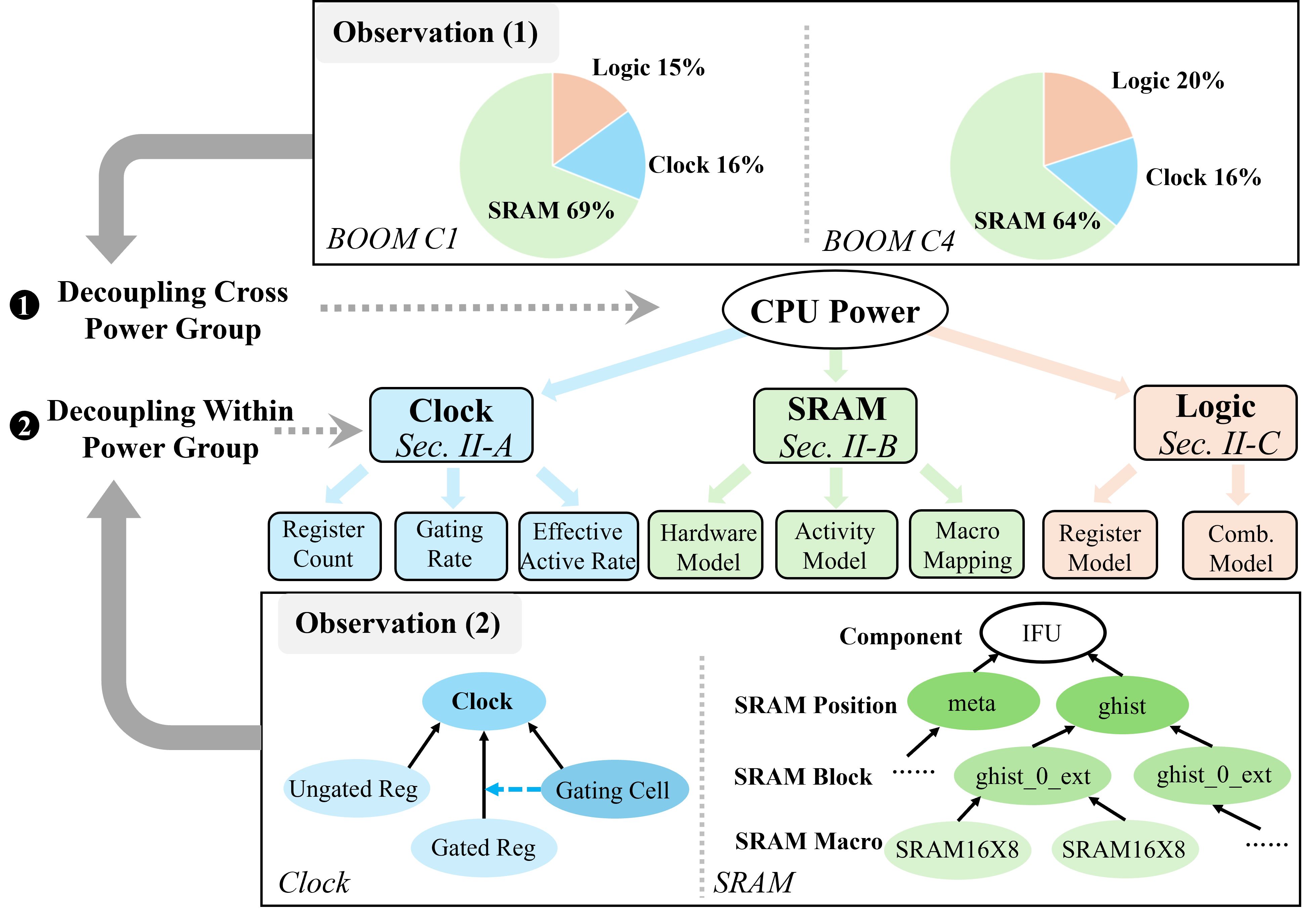}
\vspace{-.1in}
\caption{The framework of AutoPower. We have two key observations. (1) Clock and SRAM dominate the power consumption. (2) Clock and SRAM power correlate with structural information available at the architecture level. Based on these observations, AutoPower decouples the power modeling across different power groups and further decouples the power model for each power group into multiple sub-models.}
\vspace{-.2in}
\label{paradigm}
\end{figure}

\textbf{Goal:} This work proposes our new solution named AutoPower\footnote{It is open-sourced at https://github.com/hkust-zhiyao/AutoPower}, achieving both \emph{fully automatic} and \emph{few-shot} architecture-level power modeling,  
1) \emph{Fully automatic}: The power modeling method does not require any engineer-defined design-specific analytical power model. 2) \emph{Few-shot}: The model can achieve high accuracy with very limited known configurations for training.
As illustrated in Fig.~\ref{paradigm}. Our solution is initially based on two key observations: 

\begin{itemize} 
\item \textbf{Observation 1: Clock and SRAM dominate total power consumption.} The clock and SRAM dominate the power consumption of the processor. Observation 1 in Fig.~\ref{paradigm} shows power percentage of each power group of RISC-V BOOM CPU~\cite{zhao2020sonicboom} measured at layout stage. It demonstrates that the clock and SRAM dominate the power consumption of the processor. 
\item \textbf{Observation 2: Clock and SRAM power correlate with structural information available at the architecture level.}. 
(1) Clock power mainly depends on clock pins in registers and clock gating information. The number of registers and the percentage of registers with clock gating are predictable even at the architecture level.
(2) SRAM in processors usually follows a four-level hierarchy from top to bottom: component, SRAM Position, SRAM Block, and SRAM Macro. Many design components have some SRAM Positions that will be filled by SRAM-dominated structures (e.g., tables and queues). Each SRAM Position consists of multiple SRAM Blocks that will be defined in RTL implementations. SRAM Block is built up with SRAM Macros supported by technology node library (i.e., SRAM compiler). 
\end{itemize}

\textbf{Power Group Decoupling in AutoPower:} (1) Motivated by our $1^\text{st}$ observation, AutoPower decouples the power models across power groups, building individual power modeling methods for different power groups including clock, SRAM, and logic. (2) Motivated by our $2^\text{nd}$ observation, within each power group, AutoPower further decouples the modeling into multiple sub-models based on our observed unique pattern of each power group.

AutoPower adopts innovative power modeling methods for each power group as shown in Fig.~\ref{paradigm}. (1) For the clock, AutoPower is the first architectural power model that explicitly considers \emph{clock gating}, an indispensable optimization that significantly reduces clock power. Based on the structure of clock power, it proposes to decouple the clock power modeling into the prediction of register count, gating rate, and our pre-defined effective active rate.
(2) For SRAM, AutoPower estimates the power based on our proposed hierarchy with a top-down approach. AutoPower transfers the feature from components to SRAM Position, then applies the hardware model and activity model to estimate hardware information and read/write frequency at the SRAM Block level, and finally combines SRAM Macro hardware information and read/write frequency with macro-level mapping.
(3) For the logic power, AutoPower decouples it into register power (without clock pins) and combinational logic power, with different modeling methods.

\textbf{Contributions}: Our contributions are summarized below. 
\begin{itemize} 
\item We propose a new architecture-level power modeling framework, AutoPower, that decouples the power model across power groups and within each power group. It requires no design-specific engineer expertise and achieves high accuracy with only a few training samples (i.e., few-shot learning).
\item AutoPower proposes innovative power models for each power group: 
1) For clock power, AutoPower considers the clock gating technology and decouples the model based on the structure of clock power. 2) For SRAM power, AutoPower estimates the SRAM Macro information and read/write frequency based on our proposed four-level hierarchy. 3) For logic power, AutoPower adopts decoupled power for combinational logic and register power respectively. 
\item We evaluate AutoPower based on 15 out-of-order RISC-V CPUs with different configurations and 8 workloads. AutoPower achieves a low MAPE of 4.36\% and a high $R^2$ of 0.96 on average only using two known configurations for training. It achieves 5\% lower MAPE and 0.09 higher $R^2$ compared with representative prior works. 
\end{itemize}

\section{Methodology}

AutoPower decouples the modeling into different power groups including clock power, SRAM power, and logic power. Within each power group, AutoPower further decouples the power models into multiple simple sub-models based on the specific structural information of each power group. 

As an architecture-level power model, AutoPower only takes hardware parameters and event parameters as input to calculate the power consumption of the processor. Hardware parameters denoted as $H$ are the parameters to determine the processor configurations, such as DecodeWidth and ICacheWay. Event parameters denoted as $E$ are the information collected from architecture-level performance simulators, by simulating a workload with a certain CPU configuration, for example, the number of cache misses and branch mispredictions.

We introduce our model for clock power (Sec.~\ref{clknet}), SRAM power (Sec.~\ref{sram}), and remaining logic power (Sec.~\ref{logic}) in this section.

\subsection{Power Model for Clock}
\label{clknet}

Clock consumes a significant percentage of power, as shown in Fig.~\ref{paradigm}. We devote a customized power modeling method for the clock-related power. Such clock-related power is dominated by the internal power of the clock pins in registers. Clock gating is one of the most effective techniques for optimizing clock power, utilizing the clock gating cells to gate the clock of the register. Therefore, AutoPower takes the effect of clock gating into account.

\begin{figure}[!t]
\centering
\vspace{-.2in}
\includegraphics[width=0.4\textwidth]{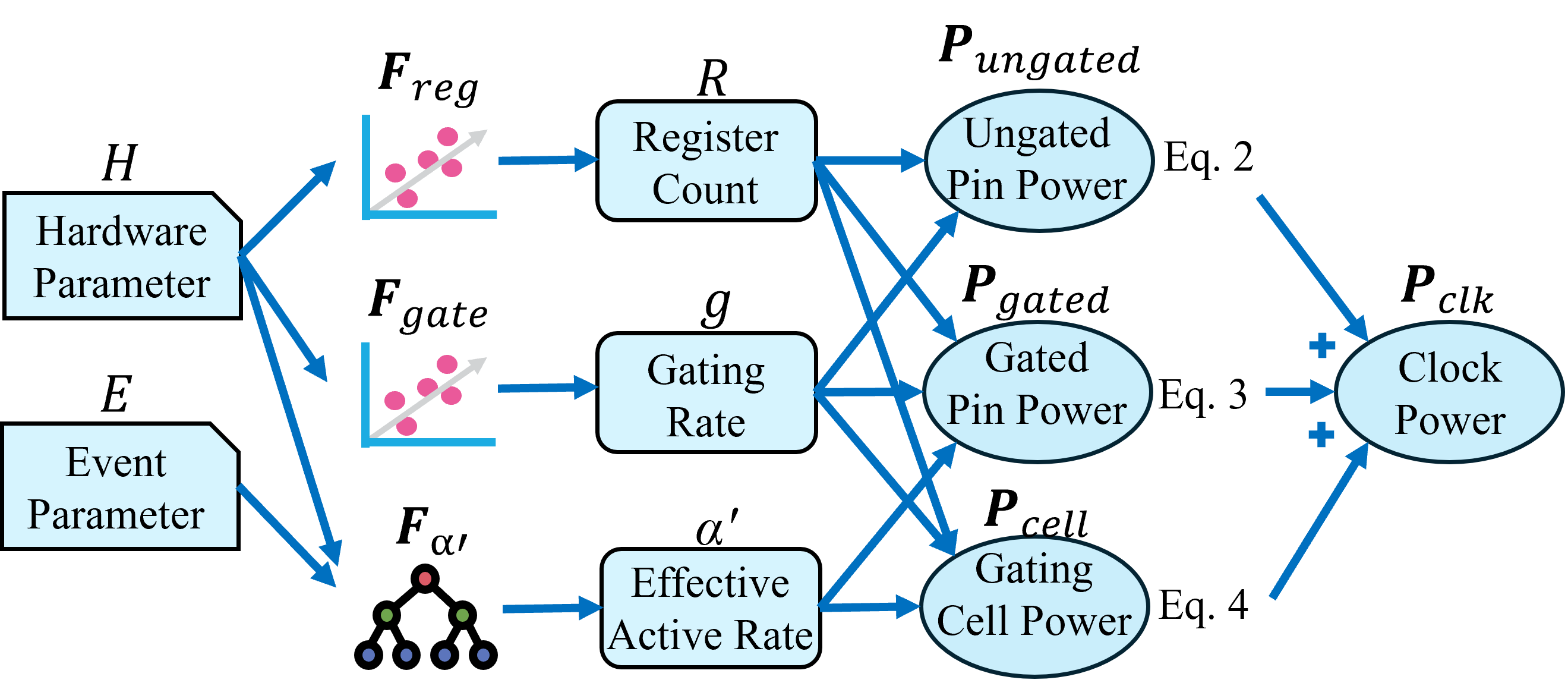}
\vspace{-.05in}
\caption{The clock power includes the power of clock pins in the ungated register (Ungated Pin Power), the power of clock pins in the gated register (Gated Pin Power), and the power of clock gating cells (Gating Cell Power). These three parts can be modeled with the register count, gating rate, and effective active rate, which are estimated by ML models.}
\vspace{-.2in}
\label{clockmodel}
\end{figure}

Fig.~\ref{clockmodel} illustrates our clock power model, which decouples the clock power into the power of clock pins in the ungated register (Ungated Pin Power), the power of clock pins in the gated register (Gated Pin Power), and the power of clock gating cells (Gating Cell Power). To accurately model these three parts of the clock power, we further decouple them into formulations represented by register count, gating rate, and effective active rate. Therefore, ML models only require estimating these three values, which have simpler correlations with hardware parameters and event parameters.

\textbf{Formulation of Clock Power:}
The clock power of a processor with the clock gating technique consists of the register clock pin power and the power of clock gating cells. The register clock pin power includes the power of clock pins in the ungated register (ungated pin power) and the power of clock pins in the gated register (gated pin power). Denoting the clock power as $P_\text{clk}$, ungated pin power as $P_\text{ungated}$, gated pin power as $P_\text{gated}$, and the power of clock gating cell as $P_\text{cell}$, the clock power can be formulated below, 
\begin{align}
P_\text{clk} = P_\text{ungated} + P_\text{gated} + P_\text{cell} \label{eq:clk}
\end{align}

We further decouple the power of each part. To formulate it, we denote the number of registers as $R$, the ratio between registers and clock gating cells as $r$, the percentage of gated registers as $g$, the clock pin power per register with active clock as $p_\text{reg}$, the pin power of latch in clock gating cell as $p_\text{latch}$, and the average active rate across all gated registers as $\alpha$. We formulate $P_\text{ungated}$, $P_\text{gated}$, and $P_\text{cell}$ below,
\begin{align}
P_\text{ungated} =& \, R * (1 - g) * p_\text{reg} \label{eq:ungated}\\
P_\text{gated} =& \, \alpha * R * g * p_\text{reg} \label{eq:gated}\\
P_\text{cell} =& \, r * R * g * p_\text{latch} \label{eq:cell}
\end{align}

(1) For the ungated pin power $P_\text{ungated}$ in Eq.~\ref{eq:ungated}, the clock pin keeps active for every cycle, regardless of the workload that is executed on the processor. Therefore, the power is the multiplication between the number of ungated registers $R * (1-g)$ and the clock pin power per register with active clock $p_\text{reg}$. (2) For the gated pin power in Eq.~\ref{eq:gated}, because of the clock gating, the clock pin can be inactive for some cycles. Therefore, the active rate $\alpha$ should be considered. (3) For the power of the clock gating cell in Eq.~\ref{eq:cell}, 
it is difficult to estimate all of these factors accurately at the architecture level. However, in processors, the number of clock gating cells is approximately proportional to the number of gated registers. Besides, the negligible effect of the toggling of its control signal and the number of registers connected with its output clock can be ignored under the architecture-level power modeling scenarios. Therefore, the power can be formulated as Eq.~\ref{eq:cell}.

Putting them together, Eq.~\ref{eq:clk} can be expressed as below,
\begin{align}
P_\text{clk} =& R\, (1 - g) \, p_\text{reg} + \alpha R\, g \, p_\text{reg} + r\, R \, g \, p_\text{latch} \nonumber \\
=& R \, (1 - g) \, p_\text{reg} + \alpha \, R \, g \, p_\text{reg} \, (1 + r \, p_\text{latch} \, /\, p_\text{reg})
\label{eq:clktogether}
\end{align}

The reason why we do not directly estimate the number of ungated registers and the number of gated registers is that the clock gating is performed with the logic synthesis tools, which is difficult to predict at the architecture level. In contrast, estimating the total number of registers $R$ and the gating rate $g$ can be easier.

To facilitate the ML-based prediction, we further simplify $\alpha * (1 + r p_\text{latch} / p_\text{reg})$ as an effective active rate $\alpha'$,
\begin{align}
\alpha' =& \alpha * (1 + r \, p_\text{latch} \, / \, p_\text{reg}) 
\end{align}
then Eq.~\ref{eq:clktogether} can be finalized as below,
\begin{align}
P_\text{clk} =& R \, (1 - g) \, p_\text{reg} + \alpha' R \, g \label{eq:clkfinal}
\end{align}
In Eq.~\ref{eq:clkfinal}, $P_\text{reg}$ can be looked up from the library file of the technology node adopted for the VLSI flow. Therefore, we totally require three prediction models. (1) register count prediction model $\boldsymbol{F_\text{reg}}$ to predict $R$, (2) gating rate prediction model $\boldsymbol{F_\text{gate}}$ to predict $g$, and (3) effective active rate prediction model $\boldsymbol{F_{\alpha'}}$ to predict $\alpha'$.

\textbf{Prediction Model:}
We formulate the three prediction models below,
\begin{align}
R = \boldsymbol{F_\text{reg}}(H), \ \  g = \boldsymbol{F_\text{gate}}(H), \ \ \alpha' = \boldsymbol{F_{\alpha'}}(H,E)
\end{align}

Register count and gating rate of a processor are determined after logic synthesis regardless of workload executed. Therefore, register count prediction model and gating rate prediction model only take hardware parameters of the component as input features. We adopt ML models to learn the correlation. Because the correlation is usually simple and we do not have sufficient samples for training, we adopt the linear model with L2 normalization as our ML model. To collect labels for training, we can collect the number of registers and the number of gated registers from the netlists of known configurations.

Different from register count and gating rate, the active rate $\alpha$ depends on both configuration and workload executed, which is the same case for the effective active rate $\alpha'$. Therefore, the effective active rate prediction model takes both the hardware parameters and the event parameters as input features.
We also adopt the ML model to learn the correlation. In contrast to the register count and gating rate prediction where the correlation is relatively simple, this correlation can be relatively complex. Besides, samples on known configurations for training can be more than the register count and gating rate because we can get multiple samples on a configuration by executing different workloads. Therefore, we adopt a relatively advanced ML model, XGBoost~\cite{chen2016xgboost}, to capture the complex correlation. 

\subsection{Power Model for SRAM}
\label{sram}

\begin{figure*}[!t]
\centering
\vspace{-.3in}
\includegraphics[width=0.98\textwidth]{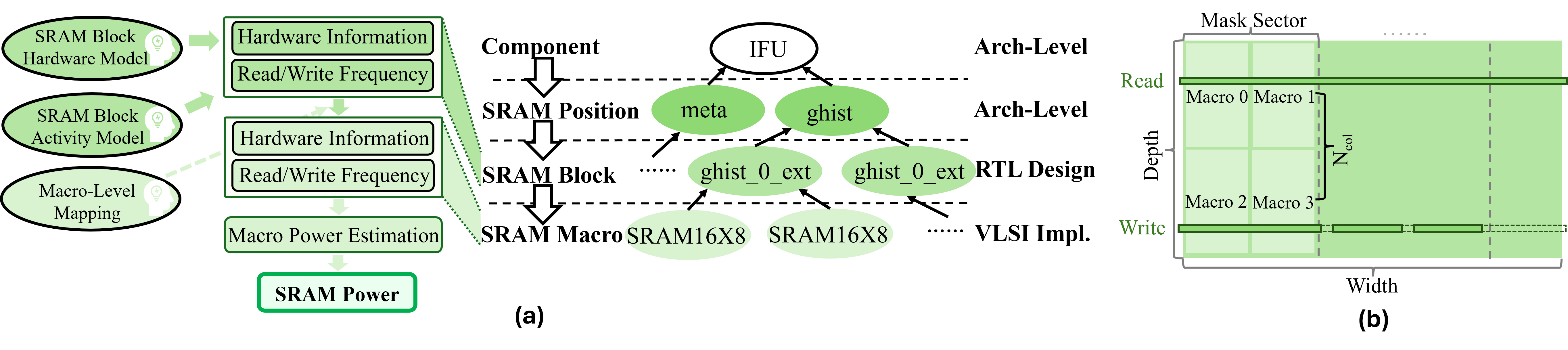}
\vspace{-.15in}
\caption{(a) SRAM power model follows a top-down four-level SRAM hierarchy of processor. For each SRAM Position in a component, SRAM Block hardware model estimates hardware information of SRAM Blocks, and SRAM Block activity model estimates read/write frequency of SRAM Blocks. Based on the estimation, macro-level mapping calculates hardware information and read/write frequency of SRAM Macros, then SRAM Power can be estimated. (b) Illustration of macro-level mapping. It shows the relation between SRAM Block read/write and SRAM Macro read/write.}
\vspace{-.2in}
\label{sramdecouple}
\end{figure*}

SRAM plays an important role in modern processors as the backbone of several tables, such as the history table in branch predictor and the metadata table in IFU, and caches. Its high power consumption is caused by expensive SRAM access. 
To estimate the SRAM power, we propose to explore the general hierarchy of SRAM in modern CPUs. 
Our model targets to estimate the SRAM power based on the SRAM Macro information, including the type, number, and read/write frequency. Then the SRAM power can be accurately estimated with the technology node library. 
Our SRAM power model mainly consists of (1) a scaling-pattern-based hardware model to estimate the number and shape of SRAM Blocks, (2) an activity model to predict the read/write frequency of SRAM Block, and (3) a macro-level mapping model to map the SRAM Block level information to SRAM Macro level.

\textbf{SRAM Hierarchy:}
As shown in Fig.~\ref{sramdecouple}(a), we propose to model the SRAM power following our proposed general four-layer hierarchy with a top-down approach:
\begin{align}
\text{Component}\to\text{SRAM Position}\to\text{SRAM Block}\to\text{SRAM Macro} \nonumber
\end{align}

SRAM Position:
At the architecture level, some components include one or more \emph{SRAM Position} that are built from SRAM and serve different functionalities. For example, in IFU, the Fetch Target Queue has two SRAM Positions: 1) ghist, the SRAM structure for the queue; 2) meta, the SRAM structure to store some other metadata.

SRAM Block: At the RTL design level, to build up an SRAM Position, one or multiple identical \emph{SRAM Block} will be used to implement a single- or multi-bank structure. SRAM Block is a logical SRAM with the shape and type required for building SRAM Position. It is the lowest-level layer that is visible to microarchitecture and independent of the support of the adopted technology node library. For example, as shown in Fig.~\ref{sramdecouple}(a), the \emph{ghist} consists of multiple SRAM Blocks called ghist\_0\_ext.

SRAM Macro: At the VLSI implementation level, \emph{SRAM Macro} is the macro supported by the adopted technology node library. Memory compiler in the technology node library cannot generate memory macros with arbitrary shapes. 
Therefore, the SRAM Block with an unsupported shape should be built up with multiple SRAM Macro supported by the memory compiler. This step is microarchitecture-invisible and is a part of the VLSI flow, where the VLSI flow usually adopts an automatic script to generate a plan to build up each input SRAM Block with the supported SRAM Macros.

\textbf{Power Model Overview:}
For the ultimate calculation of the SRAM power, we (1) transfer the feature from the Component to the SRAM Position, (2) propose a hardware model and activity model to estimate the hardware information and read/write frequency for the SRAM Block based on the feature of SRAM Position, and (3) then propose macro-level mapping to transform the SRAM Block level estimation to the SRAM Macro level for ultimate power calculation.

\textbf{Information at the SRAM Position Level:} Similar to the Component, the SRAM Position is also visible at the architecture level. Therefore, the information at the SRAM Position level includes the hardware parameter $H$ and event parameter $E$ of its component.

\textbf{SRAM Block Hardware Model:}
The hardware model takes the hardware parameter of an SRAM Position as input to estimate the hardware information at the SRAM Block level, including the width, depth, and count of SRAM Blocks to build the SRAM Position. 

The estimation is challenging because input information only includes high-level hardware parameters for SRAM Position. With our observation of SRAM scaling in processors, we propose a scaling-pattern-based hardware model for estimation. 

Our insight is that the scaling of SRAM Blocks with respect to hardware parameters $H$ mainly follows two general patterns: (1) Capacity scaling, the total SRAM capacity scales linearly with a hardware parameter, and (2) Throughput scaling, the width or number of SRAM blocks scales linearly with a hardware parameter to provide more throughput. Based on the two scaling patterns, we can automatically infer the number and shape of SRAM Block even with limited known configurations during training.

The core of our hardware model is to detect which specific hardware parameters the throughput or capacity linearly scales with and scaling coefficient. Based on the information, we can get formulations of width, depth, and count of the SRAM Block. 

\begin{table}[!t]
\centering
\renewcommand{\arraystretch}{1.1}
\resizebox{0.48\textwidth}{!}{
\begin{tabular}{c||c|c|c||c|c|c}
\hline
\multirow{2}{*}{Training Config} & \multicolumn{3}{c||}{Hardware Parameter}      & \multicolumn{3}{c}{Block Information} \\
\cline{2-7}
                                 & FetchWidth & DecodeWidth & FetchBufferEntry & width       & depth      & count      \\
\hline
C1                               & 4          & 1           & 5                & 120         & 8          & 1          \\
C15                              & 8          & 5           & 40               & 240         & 40         & 1         \\
\hline
\end{tabular}
}
\caption{The metadata table example for SRAM Block hardware model. It includes two known configurations C1 and C15 for training.}
\vspace{-.2in}
\label{hwexample}
\end{table}

To detect these hardware parameters, the model tries all hardware parameter combinations to fit a directly proportional function based on known configurations for training and selects the best combination with minimal error. 
We take the metadata table (meta) in IFU as an example. The hardware parameters of the metadata table include FetchWidth, DecodeWidth, and FetchBufferEntry. 
Table~\ref{hwexample} shows the hardware parameters and SRAM Block information of two known configurations for training. 
The hardware parameter combinations include $\{$FetchWidth$\}$, $\{$DecodeWidth$\}$, $\{$FetchBufferEntry$\}$, $\{$FetchWidth, DecodeWidth$\}$, $\{$FetchWidth, FetchBufferEntry$\}$, $\{$DecodeWidth, FetchBufferEntry$\}$, and $\{$FetchWidth, DecodeWidth, FetchBufferEntry$\}$. The model checks each combination by fitting a directly proportional function and gets its error. For example, when checking if the capacity scales with $\{$FetchWidth, DecodeWidth$\}$, the model fits the function below,
\begin{align}
\text{Capacity} = k * \text{FetchWidth} * \text{DecodeWidth} \nonumber
\end{align}
The capacities, calculated by width * depth * count, of known configurations are 120 * 8 * 1 = 1920 bit and 240 * 40 * 1 = 19200 bit, the FetchWidth * DecodeWidth are 4 * 1 = 4 and 8 * 5 = 40. Therefore, the model can get the $k$ as 240 and the error on two configurations for training as 0. This indicates the capacity scales linearly with FetchWidth and DecodeWidth, and results in the formulation of capacity as below,
\begin{align}
\text{Capacity} = 240 * \text{FetchWidth} * \text{DecodeWidth} \nonumber
\end{align}
Similarly, we can get the formulation of throughput and width,
\begin{align}
\text{Throughput} = 30 * \text{FetchWidth}, \ \ \text{Width} = 30 * \text{FetchWidth} \nonumber
\end{align}
Because count is the throughput divided by width, and depth is the capacity divided by throughput, we can get the formulation of count and depth,
\begin{align}
\text{Count} = 1 \nonumber, \ \ \text{Depth} = 8 * \text{DecodeWidth} \nonumber
\end{align}
Based on the formulation of width, depth, and count, we can estimate the SRAM Block information with only the hardware parameters.

The reason why we do not directly fit the width, depth, and count is that these three values usually do not scale linearly with hardware parameters in many SRAM Positions such as the table of ROB.

\textbf{SRAM Block Activity Model:}
The activity model takes the hardware and event parameters of the SRAM Position as input to estimate the average read/write frequency of the SRAM Blocks. 
 
Because the factors that determine the read/write frequency are complex, we utilize an ML model for the activity model. The input feature includes three parts: (1) hardware parameters of the SRAM Position, (2) event parameters of the SRAM Position, and (3) program-level features that are independent of microarchitecture, such as the number of branch instructions. All prior works do not take the program-level features into consideration. However, we find that the inaccurate performance simulator is one of the root causes of the low accuracy of the ML-based power model, therefore, some microarchitecture-independent features that are not affected by the performance simulator should be taken into account to improve the model accuracy. To collect the labels, we extract the traces of the read and write enable signals in RTL simulation to calculate the read and write frequency. The ML model we utilized is XGBoost~\cite{chen2016xgboost}.

\textbf{Macro-Level Mapping:}
Macro-level mapping takes the hardware information and average read/write frequency of the SRAM Block as input to calculate the hardware information and read/write frequency of SRAM Macro. Hardware information of SRAM Macro includes the shape and count of required SRAM Macros to build SRAM Block.

The hardware mapping takes the SRAM Block shape (i.e. width and depth) as input to calculate the shape and count of SRAM Macros to build up the SRAM Blocks. The mapping rule is a part of VLSI flow such as BOOM VLSI flow~\cite{zhao2020sonicboom}. It is available and unchanged for all processors implemented with the same VLSI flow. Therefore, given the estimated SRAM Block information, we can use the rule to get the shape and count of SRAM Macros.

To map the read and write frequency to the SRAM Macro level, we carefully analyze the building up of SRAM Block from SRAM Macro and result in a formulation to estimate the SRAM Macro level activity. Fig.~\ref{sramdecouple}(b) illustrates the activity of SRAM Macros when an SRAM Block read or write is triggered. When reading the SRAM Block, only the row of SRAM Macros that correspond to the address will be read. Therefore, the average read frequency of SRAM Macro $f^R_\text{macro}$ is that of SRAM Block $f^R_\text{block}$ divided by the number of SRAM Macros used to form a column denoted as $N_\text{col}$. 

Different from the reading, when writing the SRAM Block, because of the masking adopted for writing, only the SRAM Macros that correspond to a valid mask sector will be written. To get a unified formulation for the read and write, we define ``one write" to SRAM Block as a write to SRAM Block with all masks valid. For example, if an SRAM Block has two mask sectors, a write to this SRAM Block has one mask valid and another invalid, we will record it as 0.5 write. This will be taken into consideration when collecting the labels of write frequency for SRAM Block. Denoting the average read and write frequency per SRAM Macro as $f^R_\text{macro}$ and $f^W_\text{macro}$ and that per SRAM Block as $f^R_\text{block}$ and $f^W_\text{block}$, based on the analysis in Fig.~\ref{sramdecouple}(b), the macro-level mapping can be formulated below,
\begin{align}
f^R_\text{macro} = f^R_\text{block} / N_\text{col}, \ \ f^W_\text{macro} = f^W_\text{block} / N_\text{col}
\end{align}

\textbf{Ultimate SRAM Power Calculated at the Macro Level:}
With the number and types of SRAM Macros and the read and write frequency, we can estimate the SRAM power given the read and write power, denoted as $P_\text{R}$ and $P_\text{W}$ looked up from the technology node library. The toggling of address and data pins also has power consumption, but it only has a small contribution to the SRAM power compared with the read and write. Therefore, it can be approximately regarded as a constant.
Denoting the SRAM power as $P_\text{SRAM}$ and power caused by the toggling of address and data pins as $C$, we formulate the SRAM power as below,
\begin{align}
P_\text{SRAM} = f^R_\text{macro} * P_\text{R} + f^W_\text{macro} * P_\text{W} + C
\end{align}
where $C$ can be estimated based on the golden power of an SRAM Block collected from power simulation.

\subsection{Power Model for Logic}
\label{logic}

AutoPower models the logic power by estimating the register power and combinational logic power individually.

\textbf{Power Model for Register:}
The register power is the power of registers excluding its clock pin. Our register power model includes a hardware model and an activity model.

In detail, hardware model $\boldsymbol{F_\text{reg}}$ takes hardware parameters $H$ as input features and golden number of registers as labels. The activity model $\boldsymbol{F_\text{act}}$ takes both hardware parameters $H$ and event parameters $E$ of the component as input features. The active rate label is $P_\text{reg} / R$. The register power is the multiplication of $\boldsymbol{F_\text{reg}}$ and $\boldsymbol{F_\text{act}}$,
\begin{align}
P_\text{reg} = \boldsymbol{F_\text{reg}}(H) * \boldsymbol{F_\text{act}}(H,E)
\end{align}

\textbf{Power Model for Combinational Logic:}
Estimating the power consumption of the combinational logic is challenging. Compared with the clock, the SRAM, and the register, the pattern of combinational logic power is more complex. Besides, the combinational logic has a variety of cell types with different power characteristics, incurring the difficulty of decoupling. 

Therefore, compared with directly decoupling along physical information, we build a combinational logic power model that decouples the power into a stable power and a variation. The stable power is the power value that can reflect the general combinational logic characteristics from the power aspect, where we adopt the average power across all workloads in the fixed benchmarks for training. It is a purely hardware-related model, only depending on hardware parameters. The variation is the ratio between combinational logic power and stable power, which represents workload-specific information. 

In detail, the stable model $\boldsymbol{F_\text{sta}}$ is trained based on the average combinational power across all workloads $P_\text{sta}$, with the hardware parameters as input features. The variation model $\boldsymbol{F_\text{var}}$ takes both hardware parameters and event parameters as the input features and the ratio between power and the stable power $P_\text{comb}/P_\text{sta}$ as labels. The combinational logic power prediction can be formulated as below,
\begin{align}
P_\text{comb} = \boldsymbol{F_\text{sta}}(H) * \boldsymbol{F_\text{var}}(H,E)
\end{align}

\section{Experimental Results}

\subsection{Experiment Setup}

\begin{table}[!t]
\centering
      \resizebox{0.5\textwidth}{!}{
        \begin{tabular}{ |c||c c c c c c c c c c c c c c c| } 
\hline
Hardware Parameter  & C1 & C2 & C3 & C4 & C5 & C6 & C7 & C8 & C9 & C10 & C11 & C12 & C13 & C14 & C15 \\
\hline
\hline
FetchWidth & 4 & 4 & 4 & 4 & 4 & 8 & 8 & 8 & 8 & 8 & 8 & 8 & 8 & 8 & 8 \\
\hline
DecodeWidth & 1 & 1 & 1 & 2 & 2 & 2 & 3 & 3 & 3 & 4 & 4 & 4 & 5 & 5 & 5 \\
\hline
FetchBufferEntry & 5 & 8 & 16 & 8 & 16 & 24 & 18 & 24 & 30 & 24 & 32 & 40 & 30 & 35 & 40 \\
\hline
RobEntry & 16 & 32 & 48 & 64 & 64 & 80 & 81 & 96 & 114 & 112 & 128 & 136 & 125 & 130 & 140 \\
\hline
IntPhyRegister & 36 & 53 & 68 & 64 & 80 & 88 & 88 & 110 & 112 & 108 & 128 & 136 & 108 & 128 & 140 \\
\hline
FpPhyRegister & 36 & 48 & 56 & 56 & 64 & 72 & 88 & 96 & 112 & 108 & 128 & 136 & 108 & 128 & 140 \\
\hline
LDQ/STQEntry & 4 & 8 & 16 & 12 & 16 & 20 & 16 & 24 & 32 & 24 & 32 & 36 & 24 & 32 & 36 \\
\hline
BranchCount & 6 & 8 & 10 & 10 & 12 & 14 & 14 & 16 & 16 & 18 & 20 & 20 & 18 & 20 & 20 \\
\hline
Mem/FpIssueWidth & 1 & 1 & 1 & 1 & 1 & 1 & 1 & 1 & 2 & 1 & 2 & 2 & 2 & 2 & 2 \\
\hline
IntIssueWidth & 1 & 1 & 1 & 1 & 2 & 2 & 2 & 3 & 3 & 4 & 4 & 4 & 5 & 5 & 5 \\
\hline
DCache/ICacheWay & 2 & 4 & 8 & 4 & 4 & 8 & 8 & 8 & 8 & 8 & 8 & 8 & 8 & 8 & 8 \\
\hline
DTLBEntry & 8 & 8 & 16 & 8 & 8 & 16 & 16 & 16 & 32 & 32 & 32 & 32 & 32 & 32 & 32 \\
\hline
MSHREntry & 2 & 2 & 4 & 2 & 2 & 4 & 4 & 4 & 4 & 4 & 4 & 8 & 8 & 8 & 8 \\
\hline
ICacheFetchBytes & 2 & 2 & 2 & 2 & 2 & 4 & 4 & 4 & 4 & 4 & 4 & 4 & 4 & 4 & 4 \\
         \hline
        \end{tabular}
        }
        \caption{The CPU configurations used in our experiment. It includes 15 BOOM CPU configurations with different scales.}
        \vspace{-.1in}
        \label{configtable}
\end{table}

\begin{table}[!t]
\centering
      \renewcommand{\arraystretch}{1.1}
      \resizebox{0.5\textwidth}{!}{
\begin{tabular}{ |c||c|c|c|c|c|c| }
\hline
Component                                                    & BPTAGE                                                             & BPBTB                                                               & BPOthers                                                                             & ICacheTagArray                                                                      & ICacheDataArray                                                      & ICacheOthers                                                                         \\
\hline
\begin{tabular}[c]{@{}l@{}}Hardware\\ Parameter\end{tabular} & \begin{tabular}[c]{@{}l@{}}FetchWidth\\ BranchCount\end{tabular}   & \begin{tabular}[c]{@{}l@{}}FetchWidth\\ BranchCount\end{tabular}    & \begin{tabular}[c]{@{}l@{}}FetchWidth\\ BranchCount\end{tabular}                     & \begin{tabular}[c]{@{}l@{}}ICacheWay\\ ICacheFetchBytes\end{tabular}                & \begin{tabular}[c]{@{}l@{}}ICacheWay\\ ICacheFetchBytes\end{tabular} & \begin{tabular}[c]{@{}l@{}}ICacheWay\\ ICacheFetchBytes\end{tabular}                 \\
\hline
Component                                                    & RNU                                                                & ROB                                                                 & Regfile                                                                              & DCacheTagArray                                                                      & DCacheDataArray                                                      & DCacheOthers                                                                         \\
\hline
\begin{tabular}[c]{@{}l@{}}Hardware\\ Parameter\end{tabular} & DecodeWidth                                                        & \begin{tabular}[c]{@{}l@{}}DecodeWidth\\ RobEntry\end{tabular}      & \begin{tabular}[c]{@{}l@{}}DecodeWidth\\ IntPhyRegister\\ FpPhyRegister\end{tabular} & \begin{tabular}[c]{@{}l@{}}DCacheWay\\ MemIssueWidth\\ DCacheTLBEntry\end{tabular}  & \begin{tabular}[c]{@{}l@{}}DCacheWay\\ MemIssueWidth\end{tabular}    & \begin{tabular}[c]{@{}l@{}}DCacheWay\\ MemIssueWidth\\ DCacheTLBEntry\end{tabular}   \\
\hline
Component                                                    & FP-ISU                                                             & Int-ISU                                                             & Mem-ISU                                                                              & I-TLB                                                                               & D-TLB                                                                & FU Pool                                                                              \\
\hline
\begin{tabular}[c]{@{}l@{}}Hardware\\ Parameter\end{tabular} & \begin{tabular}[c]{@{}l@{}}DecodeWidth\\ FpIssueWidth\end{tabular} & \begin{tabular}[c]{@{}l@{}}DecodeWidth\\ IntIssueWidth\end{tabular} & \begin{tabular}[c]{@{}l@{}}DecodeWidth\\ MemIssueWidth\end{tabular}                  & ICacheTLBEntry                                                                      & DCacheTLBEntry                                                       & \begin{tabular}[c]{@{}l@{}}MemIssueWidth\\ FpIssueWidth\\ IntIssueWidth\end{tabular} \\
\hline
Component                                                    & Other Logic                                                        & DCacheMSHR                                                          & LSU                                                                                  & IFU                                                                                 &                                                                      &                                                                                      \\
\hline
\begin{tabular}[c]{@{}l@{}}Hardware\\ Parameter\end{tabular} & All                                                                & MSHREntry                                                           & \begin{tabular}[c]{@{}l@{}}LDQEntry\\ STQEntry\\ MemIssueWidth\end{tabular}          & \begin{tabular}[c]{@{}l@{}}FetchWidth\\ DecodeWidth\\ FetchBufferEntry\end{tabular} &                                                                      &              \\     
\hline
\end{tabular}
}
\caption{Architecture-level hardware parameters for each component.}
        \label{tbl:config_event1}
        \vspace{-.2in}
\end{table}

In our experiment, we adopt Chipyard~\cite{amid2020chipyard} v1.8.1 to generate RTL code. We utilize RISC-V BOOM CPU~\cite{zhao2020sonicboom} with 15 configurations with different scales for evaluation, as shown in Table~\ref{configtable}. 
Table~\ref{tbl:config_event1} lists the hardware parameters for each component.
We have eight workloads from riscv-tests~\cite{URL:riscvtests} for evaluation, including dhrystone, median, multiply, qsort, rsort, towers, spmv, and vvadd. 

We performed RTL simulation, logic synthesis, and power simulation with Synopsys VCS\textsuperscript{\textregistered}~\cite{vcs}, Synopsis Design Compiler\textsuperscript{\textregistered}~\cite{design-compilier}, and PrimePower~\cite{ptpx} respectively. Our VLSI flow is based on TSMC 40nm standard cell library and associated Memory Compiler to generate SRAM. To collect event parameters, we utilize gem5~\cite{binkert2011gem5} as our performance simulator.

\subsection{Power Prediction Results}

\subsubsection{Summary of Baseline Methods}
We compare AutoPower with the state-of-the-art architecture-level automatic power modeling method McPAT-Calib~\cite{zhai2022mcpat} as our baseline. Besides McPAT-Calib, we also include an extra baseline for the ablation study, McPAT-Calib + Component. McPAT-Calib + Component adopts the McPAT-Calib as a building block and builds power models for each component respectively. We select XGBoost~\cite{chen2016xgboost} as the ML model adopted in McPAT-Calib and McPAT-Calib + Component, as it is the best ML model reported by McPAT-Calib~\cite{zhai2022mcpat}. 

\subsubsection{End-to-End Power Model Accuracy}
\begin{figure}[!t]
\centering
\vspace{-.25in}
\hspace{-6mm}
\subfigure[McPAT-Calib]{
    \centering
    \includegraphics[height=0.15\textwidth]{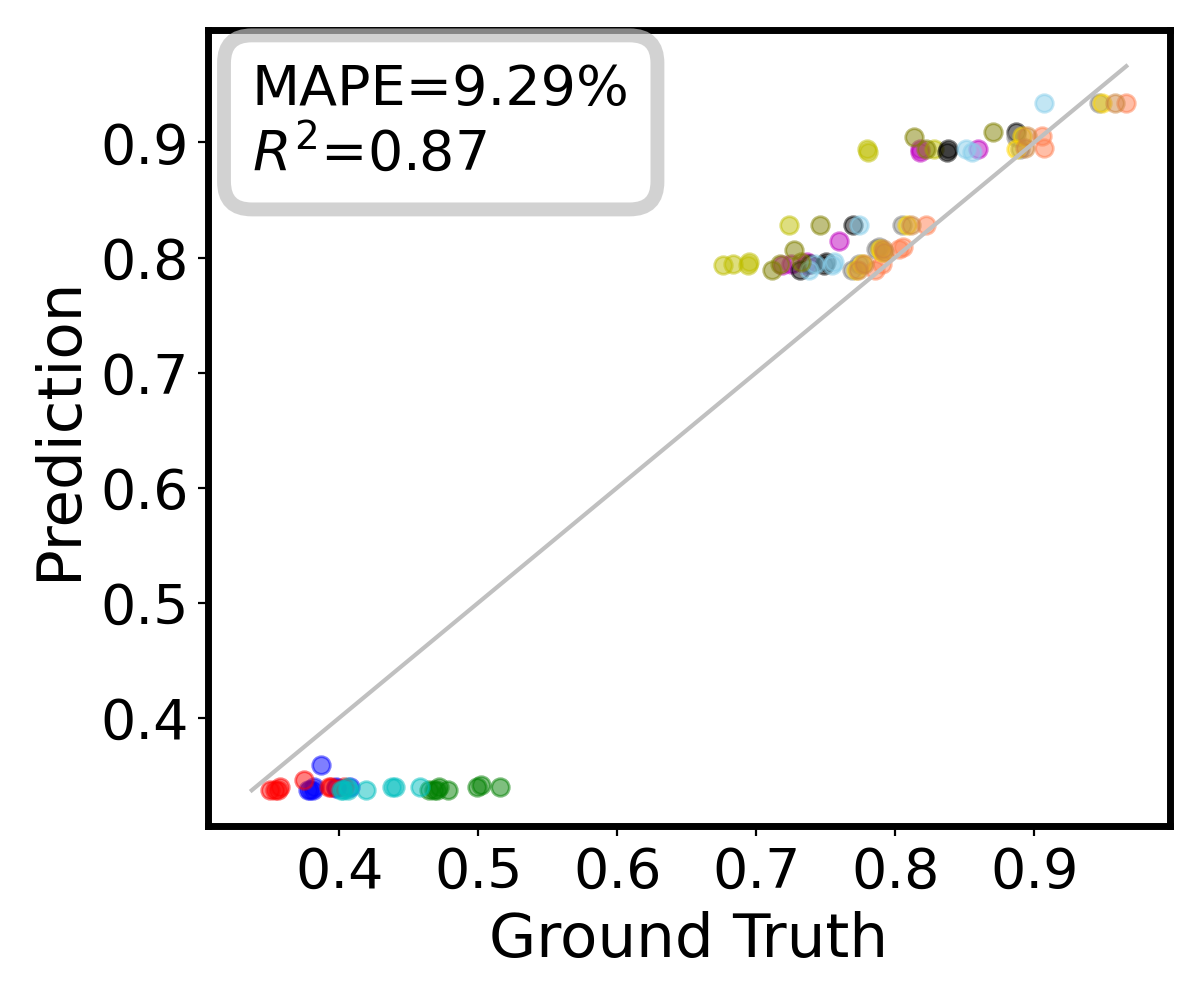}
    \label{known2_calib}
}
\hspace{-3mm}
\subfigure[AutoPower]{
    \centering
    \includegraphics[height=0.15\textwidth]{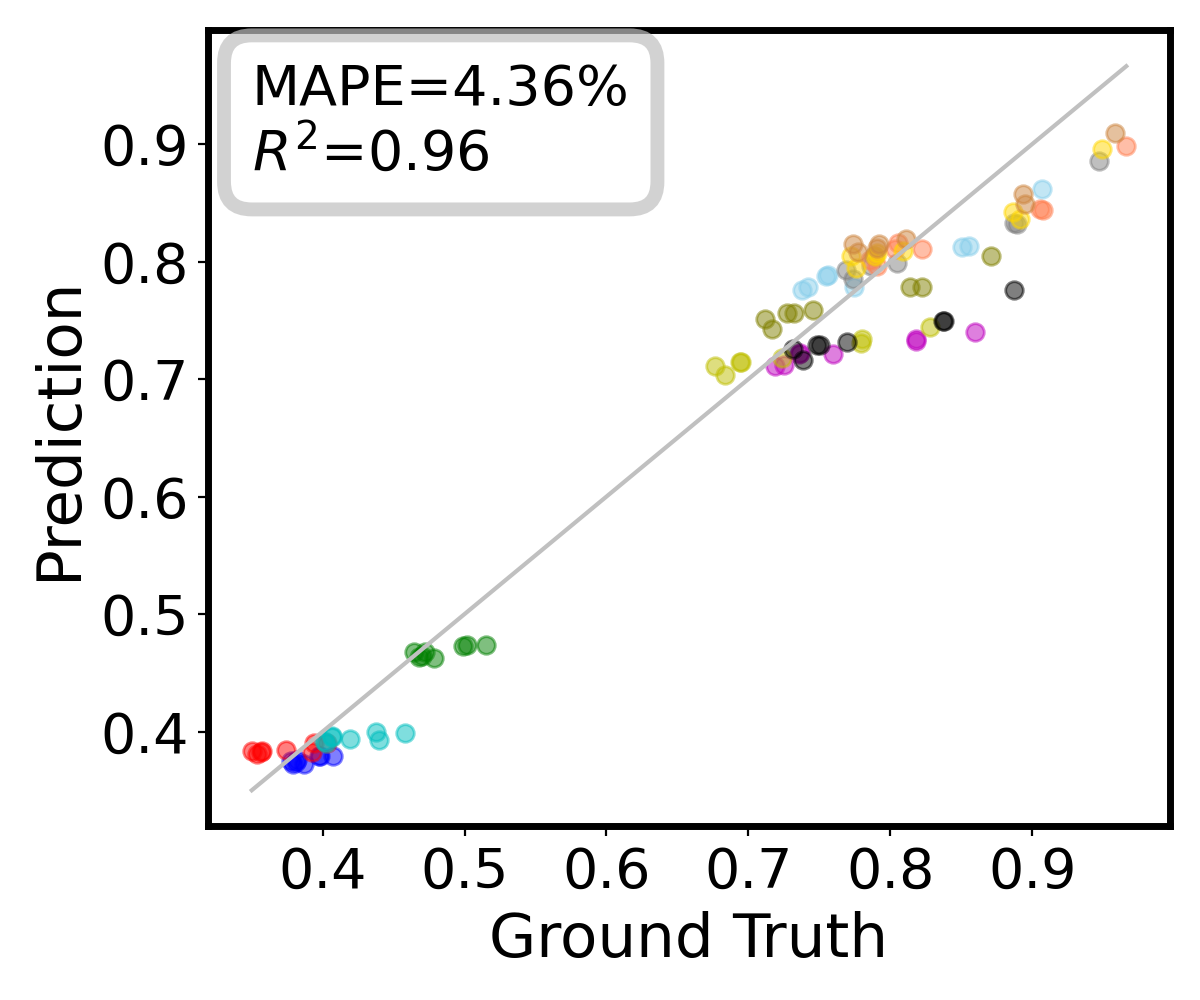}
    \label{known2_our}
}
\vspace{-.1in}
\caption{Accuracy Comparison with 2 configurations for training.}
\vspace{-.2in}
\label{known_2}
\end{figure}

\begin{figure}[!t]
\centering
\vspace{-.05in}
\hspace{-6mm}
\subfigure[McPAT-Calib]{
    \centering
    \includegraphics[height=0.15\textwidth]{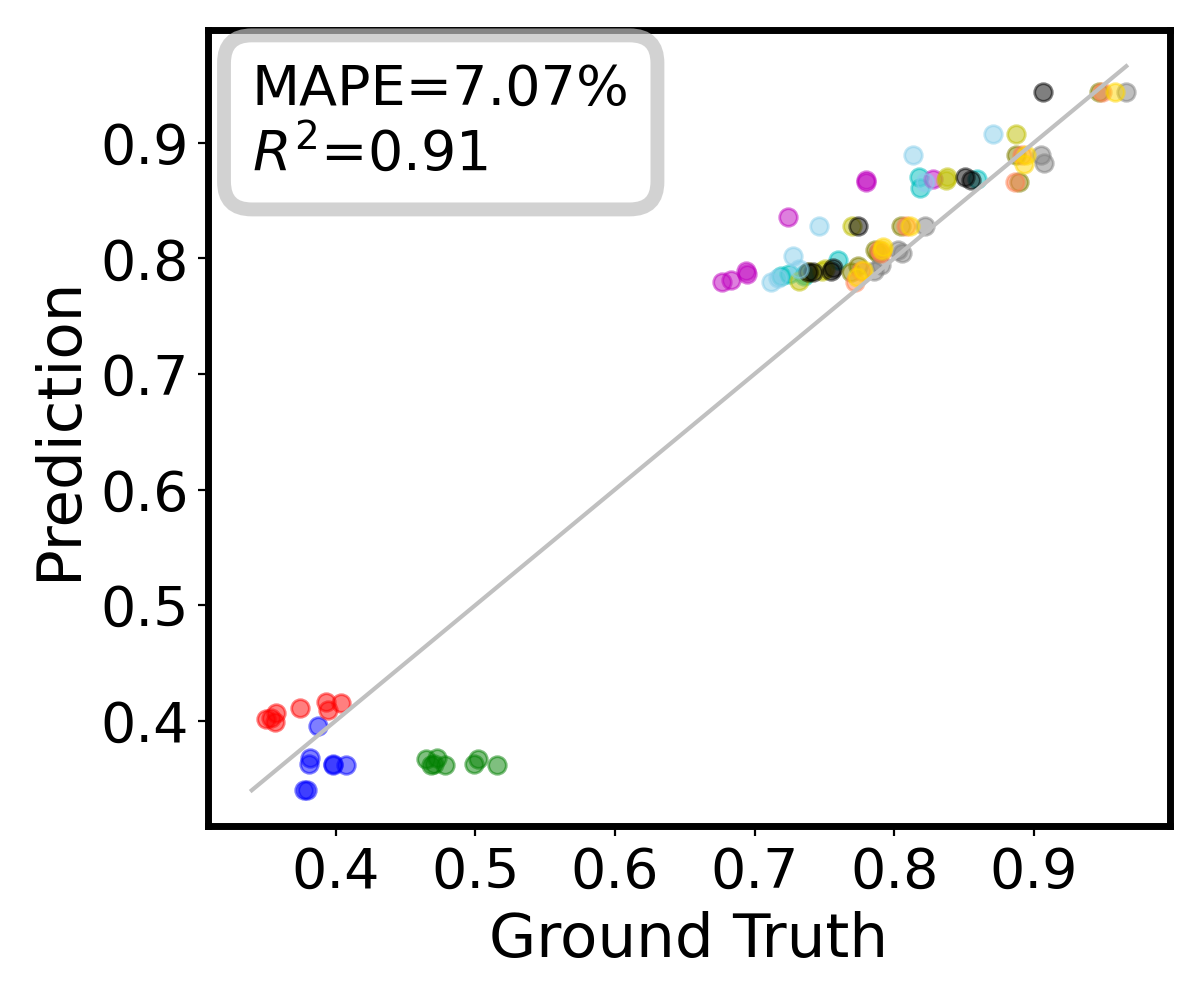}
    \label{known3_calib}
}
\hspace{-3mm}
\subfigure[AutoPower]{
    \centering
    \includegraphics[height=0.15\textwidth]{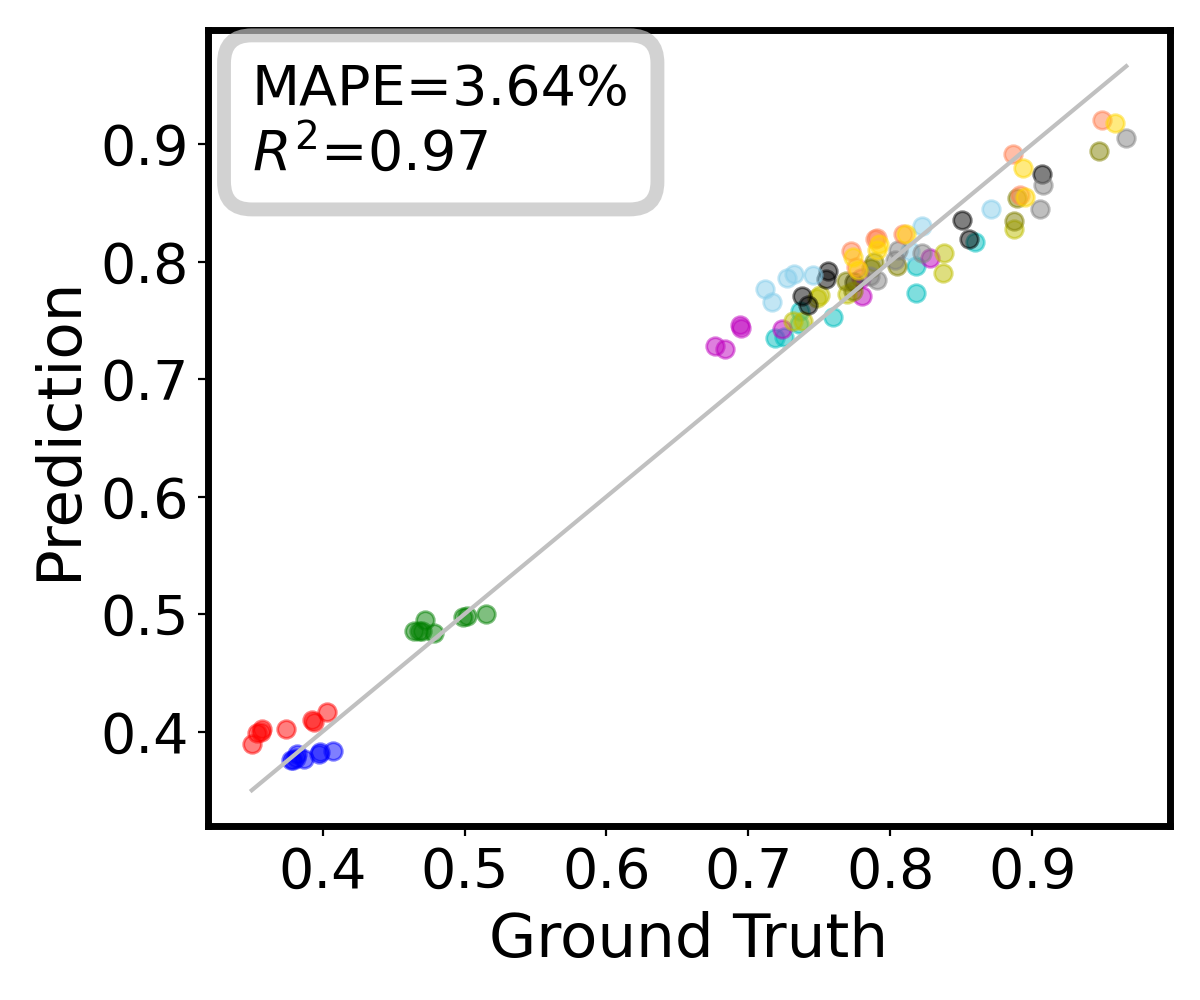}
    \label{known3_our}
}
\vspace{-.1in}
\caption{Accuracy comparison with 3 configurations for training.}
\vspace{-.1in}
\label{known_3}
\end{figure}

\begin{figure}[!t]
\centering
\includegraphics[width=0.4\textwidth]{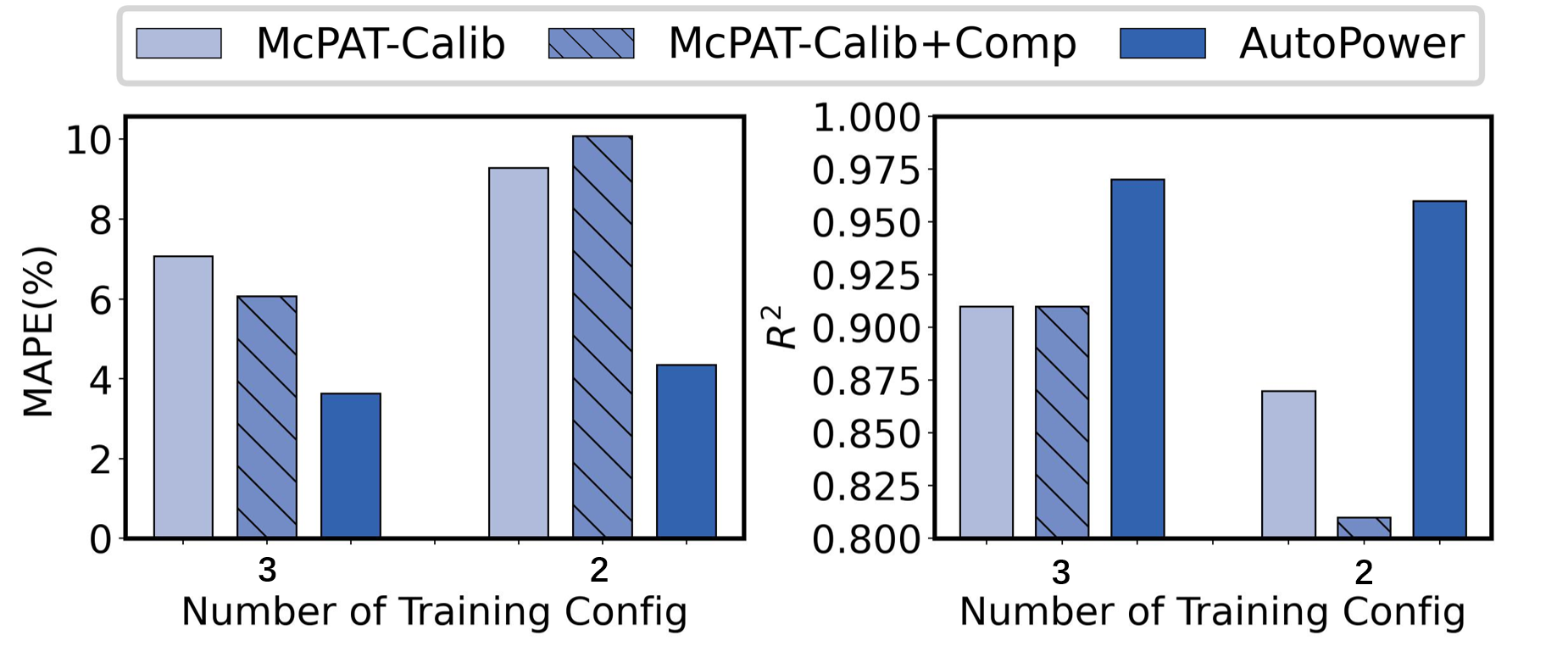}
\vspace{-.1in}
\caption{Summary of the comparison between AutoPower and
other methods under different numbers of known configurations for training. “Comp” stands for Component.}
\vspace{-.2in}
\label{ablation}
\end{figure}

Fig.~\ref{known_2} and ~\ref{known_3} visualize the comparison between AutoPower with our baseline method, McPAT-Calib, and the ablation study, McPAT-Calib + Component, when only 2 and 3 configurations are known for training when building the power model. Each point represents a configuration with a workload, points with the same configuration are in the same color. The comparisons demonstrate that AutoPower can consistently outperform McPAT-Calib. Under the two scenarios, AutoPower can achieve the highest $R^2$ of 0.96 and 0.97, which is 0.07 higher on average than McPAT-Calib with $R^2$ of 0.87 and 0.91. AutoPower can also achieve the lowest MAPE of 4.36\% and 3.64\%, which is 4.2\% lower on average than McPAT-Calib with MAPE of 9.29\% and 7.07\%. This superiority is achieved by the suitable decoupling of AutoPower. Fig.~\ref{ablation} shows the comparison results when comparing AutoPower with the extra baseline McPAT-Calib + Component. It demonstrates that AutoPower can also outperform McPAT-Calib + Component consistently. This ablation study further verifies that power group decoupling is critical for accurate power estimation.

\begin{figure}[!t]
\centering
\includegraphics[width=0.49\textwidth]{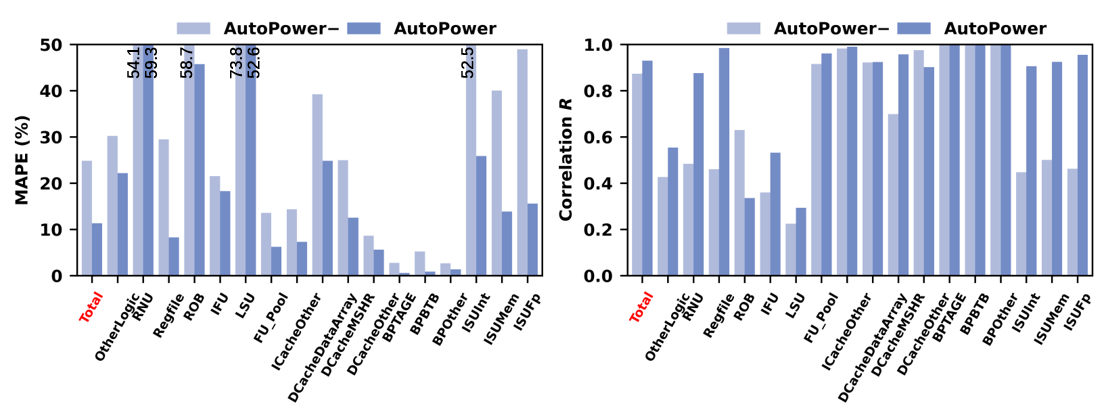}
\vspace{-.25in}
\caption{Detailed comparison of clock power prediction between AutoPower$-$ that directly utilizes the ML model and the AutoPower.}
\vspace{-.2in}
\label{clkdata}
\end{figure}

\begin{figure}[!t]
\centering
\includegraphics[width=0.49\textwidth]{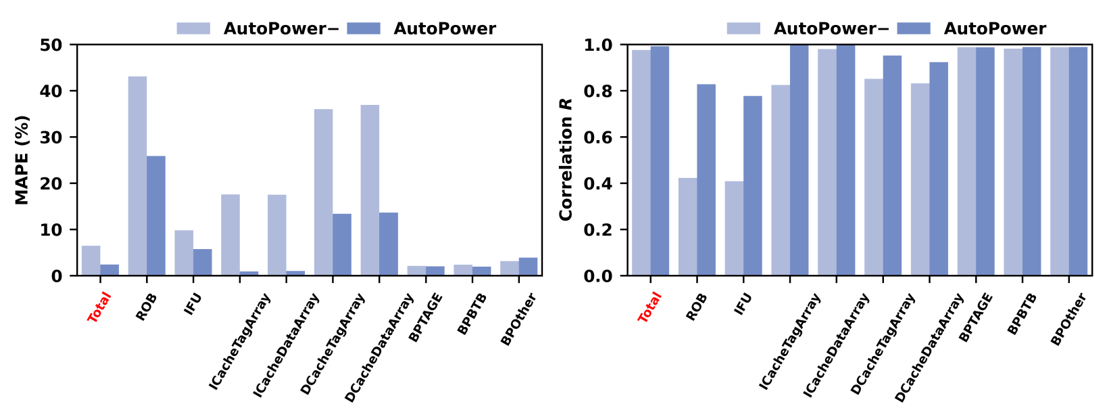}
\vspace{-.25in}
\caption{Detailed comparison of SRAM power prediction between AutoPower$-$ that directly utilizes the ML model and the AutoPower.}
\vspace{-.2in}
\label{sramdata}
\end{figure}




\subsubsection{Clock Model Accuracy}
To demonstrate the superiority of the AutoPower compared with directly applying the ML model for each power group, we provide another baseline, AutoPower$-$. It only decouples the model across different power groups and only directly adopts the ML model for the estimation of each power group.

Fig.~\ref{clkdata} shows the prediction accuracy for the clock power and the comparison with AutoPower$-$. It demonstrates that our proposed clock power model can surpass AutoPower$-$ which directly estimates the clock power for most components. Clock power prediction is challenging because of the lack of RTL-level information. 
Fig.~\ref{clkdata} also shows that, with our proposed clock power modeling method, even with limited features and limited training data, AutoPower can still achieve a relatively low MAPE of 11.37\% and a high correlation coefficient $R$ of 0.93 with 2 known configurations for training. 

The high accuracy of clock power is contributed by the accurate prediction of each sub-model. The prediction of the number of registers $R$ and gating rate $g$ can achieve a low MAPE on average with 6.93\% with 2 known configurations.

\subsubsection{SRAM Model Accuracy}

Fig.~\ref{sramdata} shows prediction accuracy for SRAM power and comparison with AutoPower$-$. The results demonstrate that SRAM power model in AutoPower can achieve a higher accuracy compared with AutoPower$-$ for most components. It can achieve high accuracy even if only the architecture-level information is available, with a low MAPE of 7.60\% and a high correlation coefficient $R$ of 0.94, under 2 known configurations. 

Such a high accuracy benefits from the accurate estimation of the number of macros and the read/write frequency. Especially, with our insight based on the scaling pattern of SRAM in processor designs, our hardware model can accurately predict the SRAM Block hardware information with a nearly 0 MAPE, and then it can accurately model the SRAM Macro with the rule from VLSI flow.

\subsubsection{Time-Based Power Trace Prediction for Large Workloads}

\begin{table}[!t]
       \vspace{-.15in}
      \centering
      \renewcommand{\arraystretch}{1.1}
      \resizebox{0.49\textwidth}{!}{
        \begin{tabular}{ c||c|c|c||c|c|c||c|c|c } 
        \hline
        Large & \multicolumn{3}{c||}{Max Power Error (\%)} & \multicolumn{3}{c||}{Min Power Error (\%)} & \multicolumn{3}{c}{Average Error (\%)} \\ 
        \cline{2-10}
        Workload & C2 & C3 & C4 & C2 & C3 & C4 & C2 & C3 & C4 \\
        \hline
        GEMM & 7.7 & 9.9 & 8.1 & 10.2 & 11.6 & 14.9 & 3.5 & 2.3 & 11.0 \\
        SPMM & 10.6 & 13.3 & 14.1 & 14.7 & 18.8 & 20.7 & 2.0 & 3.5 & 8.7 \\
        \hline
        
        \end{tabular}
        }
        \caption{Summary of fine-grained time-based power trace prediction results for large workloads with millions of cycles. It shows the percentage errors for maximal power, minimal power, and the average time-based power prediction error.}
        \label{timebased}
        \vspace{-.2in}
\end{table}

Besides average power, fine-grained time-based power trace prediction is also critical for modern processor design. However, because of the fine-grained information required for prediction, architecture-level power trace prediction is challenging.

We adopt two large workloads with millions of cycles, GEMM and SPMM, as our target workloads for time-based power trace prediction. The length of the time step is 50 cycles. The model is trained based on only two known configurations. 
Table.~\ref{timebased} shows the accuracy when applied to the C2, C3, and C4. It demonstrates that, even for fine-grained time-based power trace prediction, our model trained on two known configurations can still achieve a low MAPE without tuning on additional time-based power trace data.

\section{Conclusion}

In this work, we propose AutoPower targeting fully automated architecture-level power modeling with limited known design configurations. It decouples the modeling into different power groups and further decouples the power models
into multiple simple sub-models based on the specific structural information of each power group. AutoPower lowers the data requirement barrier, which is a compelling addition to architects’ toolbox.

\section*{Acknowledgement}
\vspace{-.02in}
This work is supported by National Natural Science Foundation of China (NSFC) 62304192, Hong Kong Research Grants Council (RGC) ECS Grant 26208723, and ACCESS – AI Chip Center for Emerging Smart Systems, sponsored by InnoHK, Hong Kong SAR. We thank HKUST Fok Ying Tung Research Institute and National Supercomputing Center in Guangzhou Nansha Sub-center for computational resources.
\vspace{-.05in}

\bibliographystyle{IEEEtran}
\bibliography{references_1, references_2, references_3}
\end{document}